\documentclass[12pt]{article}
\usepackage{amssymb,amsmath,amsfonts,multirow,rotate,color,graphicx}
\usepackage{hyperref}
\usepackage{tikz,lipsum,lmodern}

\usepackage[most]{tcolorbox}
\definecolor{gray60}{gray}{0.9}
\definecolor{gray80}{gray}{0.95}

\usepackage{graphicx}

\usepackage{times}

\usepackage{booktabs}

\usepackage[most]{tcolorbox}

\newenvironment{sciabstract}{%
\begin{quote} \bf}
{\end{quote}}

\newcounter{lastnote}

\usepackage{xcolor}
\definecolor{RoyalBlue}{HTML}{4169e1}
\definecolor{ForestGreen}{HTML}{228b22}
\definecolor{Mahogany}{rgb}{0.75, 0.25, 0.0}

\usepackage{setspace}
\usepackage{subfigure}
\usepackage{comment}

\usepackage{todonotes}

\title{Challenges and opportunities for digital twins in precision medicine: a complex systems perspective}

\author
{Manlio De Domenico $^{1,2,3\ast}$, Luca Allegri$^{1}$, Guido Caldarelli$^{4,5,6}$, \\
Valeria d'Andrea$^{1}$, Barbara Di Camillo$^{2,7,8}$,Luis M. Rocha$^{9,10}$,\\
Jordan Rozum$^{9}$, Riccardo Sbarbati$^{1}$, Francesco Zambelli$^{1}$\\
\\
\normalsize{$^{1}$Department of Physics and Astronomy ``Galileo Galilei''.} 
\normalsize{University of Padova, Padova, Italy}\\
\normalsize{$^{2}$Padua Center for Network Medicine, University of Padua.}\\
\normalsize{$^{3}$Padua Neuroscience Center, University of Padua.}\\ 
\normalsize{$^{4}$DSMN and ECLT Ca' Foscari University of Venice, Italy.}\\
\normalsize{$^{5}$Institute of Complex Systems (ISC) CNR unit Sapienza University, Rome, Italy.}\\
\normalsize{$^{6}$London Institute for Mathematical Sciences, Royal Institution, London, UK.}\\
\normalsize{$^{7}$Department of Information Engineering, University of Padua, Italy}\\
\normalsize{$^{8}$Department of Comparative Biomedicine and Food Science, University of Padua, Italy}\\
\normalsize $^{9}$ Department of Systems Science and Industrial Eng., Binghamton University, Binghamton, NY, USA. \\
\normalsize $^{10}$ Instituto Gulbenkian de Ciência, Oeiras, Portugal. \\
\normalsize{$^\ast$Corresponding author. E-mail: manlio.dedomenico@unipd.it}
}

\date{}

\begin{document} 


\baselineskip24pt

\maketitle


\begin{sciabstract} 

The adoption of digital twins (DTs) in precision medicine is increasingly viable, propelled by extensive data collection and advancements in artificial intelligence (AI), alongside traditional biomedical methodologies. However, the reliance on black-box predictive models, which utilize large datasets, presents limitations that could impede the broader application of DTs in clinical settings. We argue that hypothesis-driven generative models, particularly multiscale modeling, are essential for boosting the clinical accuracy and relevance of DTs, thereby making a significant impact on healthcare innovation. This paper explores the transformative potential of DTs in healthcare, emphasizing their capability to simulate complex, interdependent biological processes across multiple scales. By integrating generative models with extensive datasets, we propose a scenario-based modeling approach that enables the exploration of diverse therapeutic strategies, thus supporting dynamic clinical decision-making. This method not only leverages advancements in data science and big data for improving disease treatment and prevention but also incorporates insights from complex systems and network science, quantitative biology, and digital medicine, promising substantial advancements in patient care.
\end{sciabstract}


\section*{Introduction}

Precision medicine aims at delivering diagnostic, prognostic and therapeutic strategies specifically tailored to individuals by explicitly accounting for their genetic information, lifestyle and environment~\cite{mirnezami2012preparing}, all organised in a network structure\cite{barabasi2011network}. The success of this approach relies on at least two fundamental and non-trivial assumptions. The first assumption is that it is possible to predict, by means of computational, cellular, and organism models and to some level of accuracy, the response of a patient to a specific treatment. The second assumption is that it is possible to use heterogeneous data sources (multiomics, electronic health records, individual and social behavior, 
and so forth) to build massive databases with enough statistics to allow one to stratify a population with respect to characterizing and distinctive features of clinical interest~\cite{konig2017precision}.

It is not surprising that the field of precision medicine is growing~\cite{collins2015new,lu2023precision}, attracting the interest of national health systems for investments~\cite{wong2023singapore} and of scholars spanning a wide range of disciplines, from molecular biology to computer science, medicine, physics, and engineering. Nevertheless, precision medicine, with its revolutionary promises, is usually associated with clinical genomics~\cite{ashley2016towards} and multiomics~\cite{chen2013promise}, with a strong focus 
on the idea that combining those heterogeneous, multi-scale sources of data will lead to timely predictions about individual medical outcomes.
%
%
More recently, the attention shifted to the possibility of integrating such molecular data with traditional~\cite{hoffman2020biomedical,venkatesh2024health,jensen2012mining} and non-traditional \cite{correia2020mining} data sources of clinical relevance into a multiscale predictive modeling methodology known as a \textit{digital twin}, that allows for testing therapeutic strategies \emph{in-silico} with the ultimate goal of maximizing successful treatments and outcomes.

The first pioneering precursors of digital twins for personalized medicine came out in the early 2000’, proposing the idea of models of the human body for specific patients to improve clinical practices. They pointed out challenges that are still actual, such as the need of a structure for multiple sources data integration~\cite{hucka2003systems} and the importance of having solid mathematical models able to describe the system at the desired level of precision~\cite{hunter2003integration}. 
In the last years medical digital twins (MDT) have experienced a huge increase of interest, with the birth of many programmes devoted to them \cite{bjornsson2020digital, erol2020digital}. Some of the most significant successes in the field are the “artificial pancreas” or ARCHIMEDES program on diabetes~\cite{Eddy2003diab}, or mechanistic models of the heart used for cardiovascular disease monitoring and prevention~\cite{CorralAcero2020heart, coorey2021health}. Recent works emphasize the potentiality of having a comprehensive model for the human body that could help to understand what could be the possible consequences of a perturbation, for example caused by a viral infection ~\cite{laubenbacher2021using}, or the influence of specific drugs~\cite{laubenbacher2024digital, KamelBoulos2021}, on a specific patient. About the actual implementation of these models, network and complex systems are starting to be considered~\cite{bjornsson2020digital} after more than one decade from first proposals~\cite{\cite{milanesi2009trends}}, while approached based on AI and ML are widely adopted despite some limitations and critical aspects~\cite{Matheny2020}. Given the broad spectrum of definitions and applications, it is important to set the operational definition that will be adopted throughout this paper:

\begin{tcolorbox}[colback=gray80, colframe=black, boxrule=0.5pt]
    \textbf{A digital twin is an \emph{in-silico} framework which can be used to replicate a biological cell, sub-system, organ or a whole organism with a transparent predictive model of their relevant causal mechanisms which responds in the same manner to interventions.}
\end{tcolorbox}

 
Broadly speaking, a digital twin exchanges data with its real-world counterpart, synchronizing inputs and outputs, and together, they operate synergistically, with the digital twin informing, controlling, aiding, and augmenting the original system.

In fact, in the recent years the interest on digital twins exploded well beyond medicine, due to the increasing accessibility to memories, computational power and massive data gathering. Primarily utilized to simulate the intricate infrastructural configurations, they have been also applied to cities~\cite{batty2018inventing} and products~\cite{grieves2005product},  leveraging contemporary technologies such as data analytics, IoT-driven physical modeling\cite{niederer2021scaling}, machine learning and artificial intelligence. For instance, in the case of cities~\cite{caldarelli2023role}, it has been argued that it can be far more efficient to consider the emerging behaviour arising from the intricate web of relationships, processes and correlations that characterize a complex adaptive system~\cite{gell1994complex,holland2002complex,holland2006studying}, rather than reproducing a mere copy of it.

The comparison with current methodological advances and challenges in other fields, allows us to highlight the existing challenges in the case of precision medicine. Despite promising opportunities to create a digital copy of every individual to allow for personalized analysis and test individual-specific therapeutic strategies~\cite{laubenbacher2024digital,kamel2021digital,bjornsson2020digital} there are some caveats that might be considered. 

On the one hand, if digital twins must be designed to be a perfect replica of an individual, then the amount of required data vastly overcomes our present, and even the future, possibilities. The gigantic number of intervening functional units, from biomolecules to cells, makes any analytical or computational approach impossible. Even in the ideal case that a perfectly functioning computational  framework was technologically accessible, the nonlinear dynamics of interacting biological units leads to emergent phenomena that cannot be simply simulated or predicted, a landmark feature of complex systems~\cite{artime2022origin,rosen1987complex}. In fact, recent advances in predictive biology are based on building models of increasing complexity to reproduce the most salient characteristics of complex biological processes in engineered and natural populations~\cite{lopatkin2020predictive}. 

On the other hand, human patients have their own dynamical response to internal dysfunctions or differentiated coupling to the environment, including individual histories of host-microbiome and host-pathogen interactions, that might jeopardize any predictive model. 
Even more widely, the full individual exposome includes all past exposure to specific multi-scale environmental factors, such as diet and reactions to stressful biochemical or social conditions~\cite{vermeulen2020exposome,vineis2020new}. While the causal mechanisms in multiomic regulation can be partially reconstructed and accounted for, the full individual exposome is almost impossible to replicate or reproduce with a digital twin.

We have made great strides in capturing the exposome via collection of new types of data, such as mobile devices~\cite{steinhubl2015emerging} and social media \cite{correia2020mining}, 
However, even in the most ideal cases, unknown factors such as the level of disease progression and unmeasured lifestyle changes can lead to a broad set of distinct outcomes that make the design of digital twins very sensitive to the quantity and accuracy of input data. This technology may struggle to adapt and accurately predict these dynamic changes, leading to sub-optimal personalized treatment recommendations.

\begin{figure} [ht!]
\begin{center}
\includegraphics[angle=0, width=0.8\textwidth]{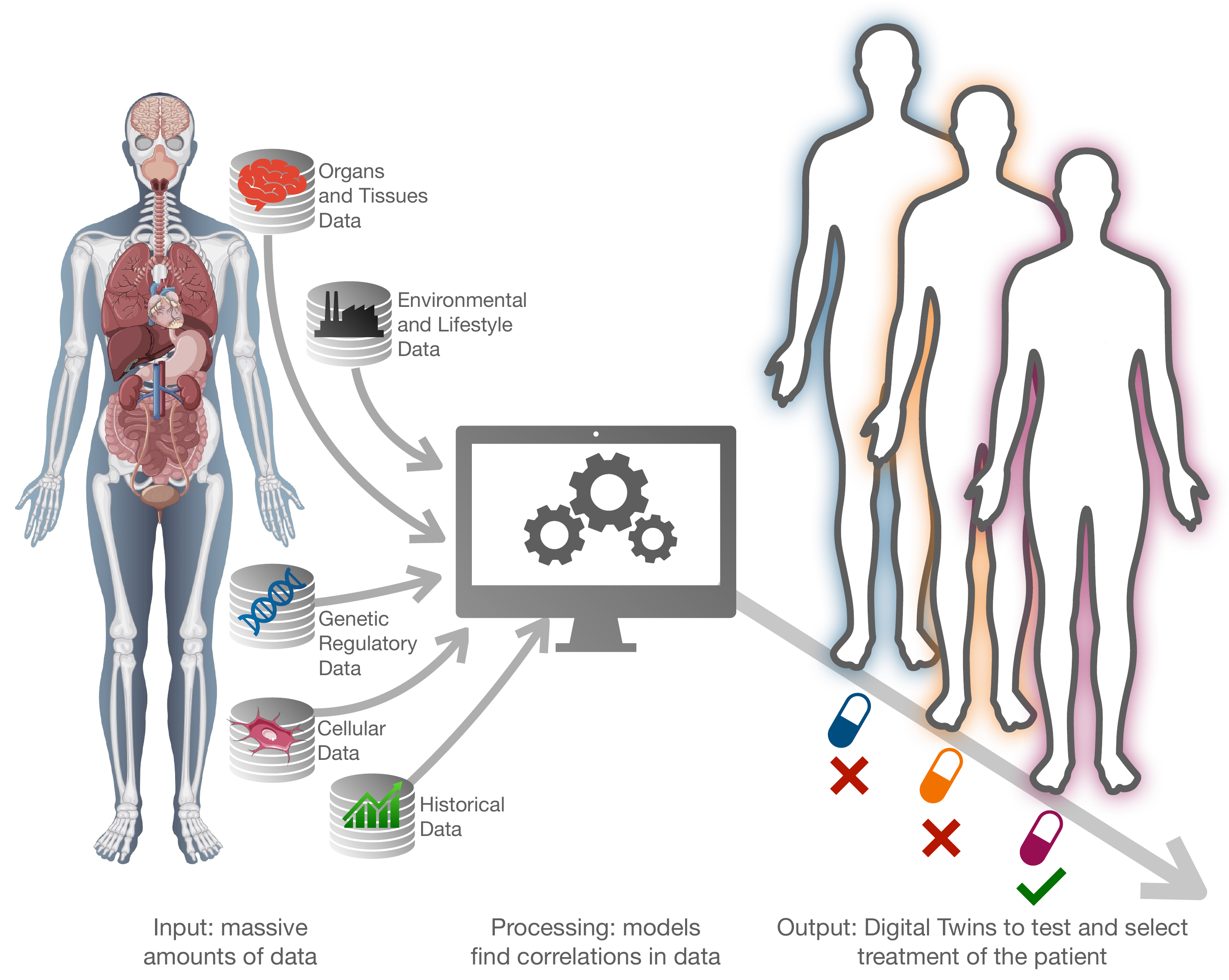}
\end{center}
\caption{
{\bf Precision medicine standard approach for digital twins.} The framework relies on using large-scale heterogeneous data sources (pre-clinical, clinical, environmental, lifestyle, etc.). This massive database should be used by sophisticated computational models (such as deep learning), while relying solely on statistical data analysis to construct a series of digitalized instances -- the digital twins -- of a patient, which will then be used to test one or more therapeutic strategies for clinical decision-making. Human body design by Freepick and osteocytes from Servier Medical Art under CC-BY AS license.}
\label{fig:fig1}
\end{figure}

The aforementioned potential issues can dramatically hinder the purpose of digital twins, suggesting that only methods based on advanced statistical data analysis, such as the one based on machine learning, are viable. However, this is not the case, since such methods provide predictive models that (i) do not easily generalize to situations and conditions for which they have not been trained for, and (ii) might recommend solutions that are clinically sub-optimal when retrieving multiple outcomes which are similarly ranked by the algorithm (Fig.~\ref{fig:fig1}).

Therefore, a more comprehensive  approach based on methods capturing the essential features of complex interconnected and interdependent systems~\cite{buldyrev2010catastrophic,de2023more} at many scales is needed. This approach needs to: (i) reduce the dimensionality of the problem of interest, by identifying the key biological, clinical and environmental variables to use for an adequate description on short time scales; (ii) characterize under which conditions a complex adaptive system like the human body (or even a cell line) can be simulated by a digital twin in terms of separated components and/or sub-systems; (iii) provide a transparent computational framework for testing actionable intervention strategies based on what-if scenarios and clinically relevant, model-informed, data-driven and evidence-based questions.

In short, this calls for more holistic and quantitative approach based on the complex adaptive nature of every patient rather than a mere replica of their salient aspects for statistical analysis.

\section*{Multiscale modeling in health and disease: from genes to systems}

Accounting for the multiscale nature of biological systems is of paramount importance for designing effective digital twins. The recent progresses in study of complex systems, especially the ones with interconnected and interdependent structure, dynamics and function, provide a promising ground for figuring out and illustrating how diverse functional units and sub-systems interact at different scales. 
Indeed, in addition to extracting multiscale molecular details from large omics datasets (e.g., transcriptomic, genomic, metabolomic, and microbiomic), we can now extract large-scale human behavior data of biomedical relevance from social media, mobile devices and electronic health records, including new patient-stratification principles and unknown disease correlations \cite{correia2020mining,steinhubl2015emerging,hoffman2020biomedical,sanchez2024prevalence,dolley2018big,jensen2012mining,nunez2024rare}. 
Accordingly, the  holistic integration and analysis of such multiscale data sources constitutes a novel opportunity to further improve personalization by including the exposome in the study of multilevel human complexity in disease \cite{dolley2018big,pescosolido2016social}, 
and used to inform more accurate models for predictive purposes in biomedicine~\cite{walpole2013multiscale,lopatkin2020predictive,nunez2024rare,correia2024conserved,kennedy2024multiscale}. 

At the lowest scale, gene regulatory networks are systems of interacting genes and their regulatory elements within a cell controlling the level of gene expression. Usually, in those networks the nodes represent genes and edges represent regulatory interactions between them, and they describe the timing, spatial distribution, and intensity of gene expression, thereby orchestrating various cellular processes such as development, differentiation, and response to environmental stimuli \cite{thattai2001intrinsic,levine2005gene,davidson2005gene,karlebach2008modelling}. A protein-protein interaction (PPI) network captures distinct types of interactions (e.g., physical contacts) between proteins in a cell. In PPI networks, nodes represent individual proteins, and edges encode interactions between them: they can be transient, like in signal transduction, or more stable, such as in the formation of protein complexes~\cite{vazquez2003global,szklarczyk2015string}. PPI networks provide insights into cellular processes, functional associations, and the modular organization of proteins, and analyzing the structure and dynamics of PPI networks helps uncover the underlying principles of cellular organization and function~\cite{han2004evidence,han2005effect,stelzl2005human,deeds2006simple,stelzl2006value,dittrich2008identifying,nepusz2012detecting,correia2024conserved}. Metabolic networks~\cite{jeong2000large,liu2013observability,nunez2024rare} map out the biochemical reactions occurring within an organism, detailing how individual metabolites are synthesized, degraded, and interconverted~\cite{guimera2005functional}. These networks are either composed of nodes representing metabolites and edges indicating the enzymatic reactions facilitating the transformation from one metabolite to another or bipartite networks where nodes are chemical species on one side and reactions on the other. In this latter representation, the web of metabolic interactions is more intricately woven, while in the former it is more straightforward. Beyond individual reactions, these networks highlight the interconnected nature of metabolic pathways, revealing redundancies, feedback loops, and regulatory mechanisms that maintain cellular homeostasis.

Intracellular networks have time-evolving states that describe which genes are active, which proteins are present (or phosphorylated, oxidized, ubiquitinated, etc.), the concentrations of metabolites, and so on. State evolution is often studied using ODE models, which can be fit to match experimental state and kinetic data~\cite{mackeySimpleMathematicalModels2016}. In many cases, the available data is insufficient to fully constrain the parameters of an ODE model, or, which is often the case, the underlying biological dynamics is of a threshold nature~\cite{gates2021effective}.
In these cases, a discrete causal model, such as Boolean networks (or multistate automata networks more generally), may be appropriate~\cite{schwab2020ConceptsBooleanNetwork,rozum2024BooleanNetworksPredictivea}. In a Boolean network, the state of each node in the intracellular network is binarized: a gene is either active or inactive, and the active form of a protein is either above some unspecified threshold of abundance or below it. The binarized states change in time according to logical (Boolean) update functions, i.e., each network node is an automaton~\cite{marques2013canalization}.
The causal effect of various interventions (e.g. drugs) can be evaluated by manipulating the states of individual nodes and observing the resulting dynamics.
Since Boolean automata can be grouped to model variables with more than two states, the approach is widely applicable to model cellular components with various levels of activation, e.g. proportion of cells that enter apoptosis in breast cancer cell lines~\cite{gomeztejedazanudo2021CellLineSpecific}.
A common application of these models is to study the effects of combinatorial drug interventions, particularly in the context of cancer~\cite{gomeztejedazanudo2021CellLineSpecific,folkesson2023SynergisticEffectsComplex}. To serve as a component of a digital twin, Boolean networks must reconcile their discrete time steps with physical time. This is often done by updating node states asynchronously according to tunable node transition rates, essentially treating the dynamics as a continuous Markov process~\cite{calzone2022ModelingSignalingPathways}. This approach has been applied, for example, to suggest personalized drug therapies for prostate cancer patients using personalized Boolean network models~\cite{montagud2022PatientspecificBooleanModels}.
This discrete dynamics approach has also been used to infer important dynamical pathways in multilayer networks, tying molecular factors (from multiomics, brain and retinal imaging data) to clinical phenotype (from patient data) in multiple sclerosis~\cite{kennedy2024multiscale}. This is an exciting avenue that allows complex regulatory dynamics to be studied on static multilayer networks obtained from heterogeneous data sources, whereby each node can integrate incoming signals differently---going well beyond the typical analysis via spreading or information dynamics on networks.

It is important to note that automata network models can typically be greatly simplified by reducing dynamically redundant interactions \cite{marques2013canalization}, due to the ubiquity of canalized dynamics in biology \cite{manicka2022effective,costa2023effective}. 
This results in scalable causal models capable of uncovering actionable interventions, conditioned on different input assumptions, in a transparent manner \cite{marques2013canalization,gates2021effective,rocha2022feasibility}.
%
Boolean networks are especially amenable to causal analysis
because they can be converted to simplified causal representations  (according to Boolean minimization criteria) \cite{marques2013canalization,gates2021effective,rozumPystablemotifsPythonLibrary2022,benesAEONPyPython2022}, 
standing in stark contrast to the black-box predictions of traditional machine learning methods or tallying the outputs of Monte Carlo simulations of large dynamical models (including non-simplified Boolean Networks.)

Thus, automata network models---whose parameters can be inferred and validated from perturbation experiment, multiomics, and exposome data---are ideal components to consider for the top level of digital twins, as they synthesize the large-scale underlying data into simplified, explainable, causal networks amenable investigate actionable interventions.
Indeed, these features show how this modeling approach directly responds to the digital twin approach needs identified in the introduction: dimensionality reduction, scalable modularity, and transparency.

Whether discrete or continuous, the dynamics of intracellular networks can be coupled with each other and with physical processes to produce whole-cell models, which attempt to describe the whole genome, proteome, and metabolome of a cell over the course of its life cycle in a fine-grained dynamical model~\cite{bhat2020WholeCellModelingSimulation}, as was first demonstrated in the human pathogen \emph{Mycoplasma genitalium}~\cite{karr2012WholeCellComputationalModel}. More recent efforts have been focused on identifying minimal genomes~\cite{rees-garbutt2020DesigningMinimalGenomes} or modeling organisms with larger genomes, such as \emph{E. Coli}~\cite{sun2021ColiWholeCellModeling}. Currently, the biomedical application of such detailed models is limited by the enormous effort required to construct them. Fortunately, it is often the case that only specific processes need to be incorporated to build a medically relevant digital twin. Narrowing the focus of the model at the cellular level makes model construction and personalization more feasible, lowers computational barriers, and facilitates embedding these models into multicellular models, e.g.~\cite{poncedeleon2023PhysiBoSSSustainableIntegration}.


An interesting focus is given by single cell data analysis. Indeed, tissues are complex multi-agent systems made of multiple subpopulations of cells that, even of the same type, exhibit different system state and expression profiles and that are spatially and temporally organized and able to communicate and interact with each other and to orchestrate self-assembly and response to stimuli as a whole. This is fundamental in many biological contexts, such as early embryonic development and tumor etiology, where different cells are characterized by distinctive genetic mutations and/or expression profiles. These differences are regulated by cell-to-cell communication and underlie complex dynamic responses characterizing healthy and pathological tissue development \cite{shalek2014single}.
An example of how the interaction between cells can be modelled to describe emergent behaviors is provided by agent-based models applied to study the interaction dynamics between immune and tumor cells in human cancer. Coupling a discrete agent-based model with a continuous partial differential equations-based model, these models capture essential components of the tumor microenvironment and are able to reproduce its main characteristics. Each tumor is characterized by a specific and unique tumor microenvironment, emphasizing the need for specialized and personalized studies of each cancer scenario. Recently, a model of colon cancer has been proposed  that can be informed with patient transcriptomic data \cite{cesaro2022mast}. It would be interesting to extend the model by informing it through methods that infer cellular communication \cite{baruzzo2022identify, kumar2018analysis, puram2017single, zhang2019landscape}. This would have the advantage of characterizing the tumor environment more specifically by defining the probability of an agent's action in response to received communication.

At the tissue scale, neural, cardiovascular, and respiratory systems are a few examples of systems that need to be considered. However, system dysfunctions, such as cancer, should also be incorporated. 
Several studies have focused on examining neural connections within an organism's brain, commonly referred to as the connectome \cite{Sporns2005}. Whether investigating the intricate networks within the human brain or the simpler wiring maps of organisms like \emph{C. elegans} or \emph{Drosophila Melanogaster}, the objective remains consistent: elucidating the interconnections and organization of neurons and regions at the mesoscale. Functional imaging techniques are utilized to explore the relationship between activity in specific brain areas \cite{bullmore_complex_2009}. The analytical and computational tools from network theory allow to build maps of structural and functional connections, revealing characteristics of complex networks—such as small-world topology, highly connected hubs, and modularity—that manifest at both the whole-brain and cellular scales \cite{Sporns2002, Sporns2004,Chialvo2010}. 

The human brain is an emblematic case study for the design of ambitious computational models towards digital twins. Nevertheless, despite the aforementioned significant advancements, the explicit goal of building a realistic computer simulation of the brain within a few years has not met expectations~\cite{Siva2023}.

\begin{figure} [ht!]
\begin{center}
\includegraphics[angle=0, width=\textwidth]{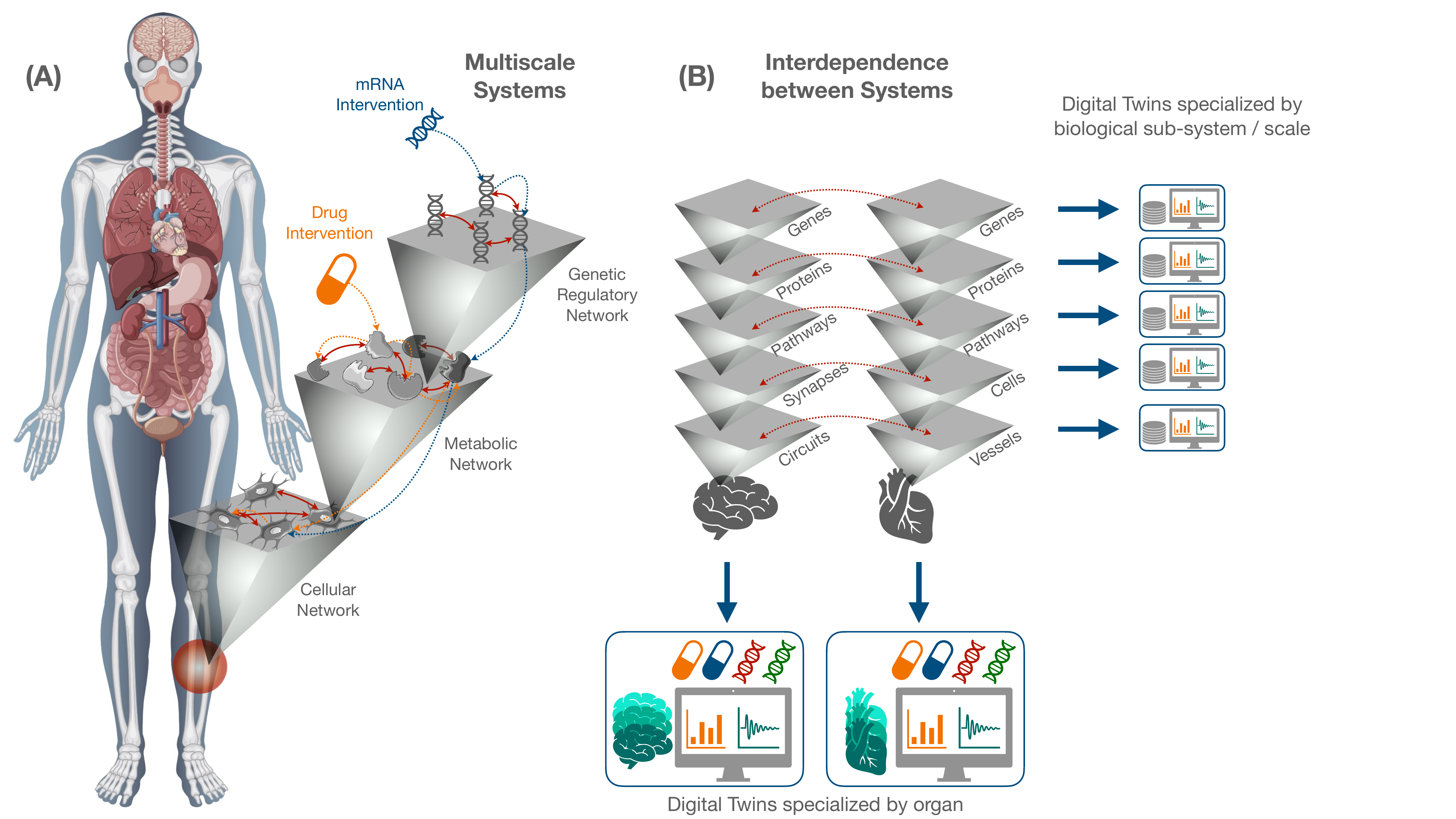}
\end{center}
\caption{
{\bf Multiscale and network modeling for digital twins.} (A) Once the potential source of a dysfunction is identified, multiple biological systems might be involved, across different spatial and temporal scales. Treatments, e.g., being based either on mRNA therapy or on classical drugs, usually target biological units at a specific scale. However, their effects might propagate to other units at the same scale or across scales due to the presence of interactions and interdependencies among biomolecules and functional sub-systems such as cells. (B) A multiscale illustration of the interdependent sub-systems related to the function of distinct organs. Each scale can be simulated by a specialized digital twin, or their multiscale integration can be simulated by a more complex, but still specialized, digital twin. The effects of distinct treatments can be analyzed on several distinct instances of the digital twins using model-informed and data-driven search. Human body design by Freepick and osteocytes from Servier Medical Art under CC-BY AS license.}
\label{fig:fig2}
\end{figure}

Overall, the aforementioned sub-systems are part of a broader complex, adaptive, interdependent system of systems which are organized in hierarchies of increasing complexity with modular organization~\cite{wagner2007road}. This is a fact well recognized since at least half a century, summarized in the Jacob's statement that “every object that biology studies is a system of systems”~\cite{jacob1974logic}. Such sub-systems exchange information (e.g., in terms of electrical, chemical and electrochemical signals) to regulate each other and operate out of equilibrium~\cite{haken1975cooperative,ritort2008nonequilibrium,fang2019nonequilibrium}.

Consequently, considering any sub-system in isolation from the other ones provides an incomplete representation of each sub-system, leading to inaccurate models and predictions of biological processes. A partial solution to this problem comes from the statistical physics of multilayer systems, allowing one to describe each scale by a \emph{level} of organization, whereas each level (i.e., at the same scale) can be characterized by multiple contexts, namely \emph{layers}~\cite{de2013mathematical,de2023more}. Levels can be interdependent with each other~\cite{buldyrev2010catastrophic,gao2012networks} while being characterized by different contexts. In the case of biological systems~\cite{de2018multilayer}, this is reflected in the distinct type of interactions among the same set of biomolecules or the distinct channels available for cell-cell communication, as well as in the interdependence between distinct systems such as the cardiovascular and nervous ones (Fig.~\ref{fig:fig2}). 

This web of interconnections and interdependencies involving diverse and heterogeneous functional biological units across scales play a pivotal role in human health, and it is plausible to associate their dysfunction to disease states~\cite{greene2017putting,halu2019multiplex}.

\section*{Challenges in multiscale modeling}

Multiscale modeling of biological systems presents formidable challenges, primarily due to the intricate and redundant networks of interactions and interdependent processes taking place and unfolding across different scales, from molecular (microscopic) to organismic (macroscopic) levels. These systems are characterized by dynamic processes that operate far from equilibrium, exchanging various types of signals -- e.g., chemical, electrochemical, and more -- thereby creating a complex ecosystem of interlinked dynamical processes~\cite{lopatkin2020predictive}. Such complexity poses significant difficulties in developing models that are both significant and coherent, avoiding extremes like reductionism, which assumes that sufficient computational power can simulate an entire organism, or oversimplification, which relies excessively on abundant data to sidestep the need for intricate modeling.

Moreover, biological systems are inherently adaptive, adjusting dynamically to environmental changes~\cite{freeman2001biocomplexity}. This adaptiveness is crucial for accurately simulating the impact of external factors such as therapeutic interventions or changes in environmental conditions like pollution~\cite{fuller2022pollution} or alterations in food sources~\cite{van2023deciphering,al2023microplastics}. Responses to these changes start at the cellular level, influencing gene and protein expressions (or lack of, that can even trigger the insurgence of cancer~\cite{Parreno2024}), and extend to higher biological structures through complex signaling pathways involving ligands and receptors. Such adaptive complexity must be integrated into models to accurately reflect the biological response to external stimuli within the spatial and temporal scale of interest.

In the broader context of precision medicine, integrating digital twins that reflect these multiscale and adaptive features poses even additional challenges. The models often employed are predominantly phenomenological, focusing more on observed phenomena rather than the underlying mechanisms, an approach resulting in a significant gap in our mechanistic understanding, which is essential for bridging various biological scales effectively. Drawing again a parallel with urban system digital twins might offer new perspectives and strategies: cities face similar multiscale integration challenges~\cite{caldarelli2023role} and require a similar framework for addressing the complex interplay of different components within a living system, potentially guiding the development of more effective biomedical models. 

By critically analyzing these challenges through the lens of complexity science, we can better understand and possibly overcome the hurdles in creating cohesive and predictive multiscale models that are crucial for the future of biomedical research and therapeutic development. In the case of interconnected systems at a given scale, one can introduce a suitable object named multilayer adjacency tensor $M^{i\alpha}_{j\beta}(t)$, to operationally encode all the interactions at time $t$ between a biological unit $i$ (e.g., a single protein or a protein complex) in a layer $\alpha$ (e.g., a class of biological processes or a pathway) and another biological unit $j$ (e.g., another protein or a metabolite). The framework is so general that it allows to include also cross-layer structural interactions, if any. In fact, due to the high number of interacting units (such as biomolecules, cells, etc), biological modeling often assumes such deterministic processes, such as that reactions occur at constant rates, compartmental interactions are fully-mixed or mean-field approximations apply. Therefore, even at some good level of approximation, the dynamics of some quantity $\mathbf{x}(t)$ of interest -- e.g., the concentration of metabolites or the population of some species (e.g., cancer cells, bacteria, etc) -- might be described by multilayer differential equations~\cite{de2016physics,de2023more} like
\begin{eqnarray}
\label{eq:model_ODE}
\frac{\partial x_{j\beta}(t)}{\partial t} = f_{j\beta}(x_{j\beta},t) + \sum_{i}\sum_{\alpha} g_{j\beta}\left[M^{i\alpha}_{j\beta}(t), x_{i\alpha}(t), x_{j\beta}(t) ,t\right],
\end{eqnarray}
where $f_{j\beta}(\cdot)$ is a function only of the variable $x_{j\beta}(t)$ corresponding to a specific unit $j$ in a specific context or layer $\beta$, whereas $g_{j\beta}(\cdot)$ is a function accounting for the interactions between pairs of units, i.e., for the effects due to the intervening networks.

It is remarkable how such a simplified deterministic framework can allow to model some response to clinical treatment triggered by basic chemical reactions, as well as that the value of PH, the actions of cells that can be triggered and even the production of proteins that can be stimulated by acting on specific part of DNA or by specific mRNA targets~\cite{kowalski2019delivering,rohner2022unlocking}. In the light of these simple arguments, it might be tempting to rely only on such deterministic approaches -- based on sets of differential equations, such as Eq.~(\ref{eq:model_ODE}), or on agent-based modeling -- to predict the behavior of a therapeutic intervention. After all, if we have systematic cause-effect relations linking interventions to biological and clinical outcomes, it would be enough to calibrate our models on the specific features of a patient to determine their response to treatments and potentially cure a disease. 

However, in complex and variable environments such as a living organism, adaptiveness, randomness and biological noise might affect the model outcomes. Adaptiveness can be still reflected by such simplified models: if we indicate with $u_{j\beta}(t)$ some external input signal or control applied to a biological system and with $\Theta$ the set of parameters that dynamically change based on the system's states or external inputs, then a more general model at a given scale could be formalized as

\begin{eqnarray}
\label{eq:model_ODE2}
\frac{\partial x_{j\beta}(t)}{\partial t} &=& f_{j\beta}(x_{j\beta},t) + \sum_{i}\sum_{\alpha} g_{j\beta}\left[M^{i\alpha}_{j\beta}(t), x_{i\alpha}(t), x_{j\beta}(t), \Theta, u_{j\beta}(t), t\right]\nonumber\\
\frac{\partial M^{i\alpha}_{j\beta}(t)}{\partial t} &=& \ell(M^{i\alpha}_{j\beta}(t), \Theta, u_{j\beta}(t), t)\nonumber\\
\frac{\partial \Theta(t)}{\partial t} &=& h(\Theta, u_{j\beta}(t), t)
\end{eqnarray}
which is much more complicated than Eq.~(\ref{eq:model_ODE}), but it can be still managed from a computational point of view. 

Noise can be inherent to one or more aspects of the involved systems -- e.g., biochemical and electrochemical variability -- or being linked to specific mechanisms altered by internal or external perturbations, such as virus-host interactions, environmental changes, so forth and so on. Accordingly, including the effects of noise depends on the scale and impact of the biological process being modeled. For instance, including DNA replication errors for the analysis of short-term effects of a therapeutic drug might not add relevant biological or clinical insights, while adding complexity to the model. Another emblematic case is the use of discretized structures, such as networks, to model processes that are manifestly continuous (e.g., in space): in such conditions, using complex networks will introduce a level of sophistication that is not necessary to gain insights about a biological process.

These noise sources introduce an additional level of stochasticity that cannot be easily taken into account by statistical models, even the most sophisticated ones based on machine learning. Nevertheless, what it usually assumed to be a bug might be a feature: as for other complex systems in nature, stochasticity is indeed structured and can lead to self-organized behaviors and processes~\cite{nicolis1977self,heylighen2001science,camazine2020self}. The theory of nonlinear dynamical systems and the statistical physics of complex networks provide suitable theoretical and computational frameworks to model such complex biological phenomena~\cite{fang2019nonequilibrium}, and should be considered as essential ingredients to design reliable digital twins, either specialized or not, for any living organism.

Nevertheless, the most important obstacle to describe realistic biological systems relies on incorporating multiple dynamic processes across the multiple intervening scales, primarily due to the diverse nature of the laws governing these processes at each scale. One significant technical challenges is effectively bridging these scales. This involves not just scaling up or down the processes, but also ensuring that interactions between scales are accurately captured. This might involve developing intermediate models or using scale-bridging techniques like homogenization or coarse-graining, which themselves can introduce approximation errors or require simplifications that might affect model accuracy. While some models are based on fundamental laws -- such as reaction-diffusion processes for chemical networks -- other models are genuinely phenomenological: reconciling dynamics of so different nature is challenging, since the latter class of models might not be suitable to capture novel phenomenology. This problem can only be partially solved by developing more fundamental models, since biological processes are characterized by emergent phenomena that cannot be directly deduced even from having full knowledge about their units and their interactions~\cite{gell1994complex,holland2002complex,holland2006studying,artime2022origin}. To this aim, one needs to account, simultaneously, for the evolution of the system according to dynamics similar to the one in Eq.~(\ref{eq:model_ODE2}) and the fact that the underlying mechanisms can change while satisfying the constraints imposed by physics and chemistry, requiring meta-dynamical models~\cite{bagley1989modeling}.

Additionally, multiscale models often require extensive parameterization, which can be difficult when experimental data are scarce at certain scales: therefore, validating these models across all scales can be exceptionally challenging, especially when direct observations or experiments at certain scales are not feasible or provide, at best, indirect measurements (such as correlations) about the phenomenon of interest that require an adequate inferential framework~\cite{peel2022statistical}.

Furthermore, such models should be able to propagate perturbations from one scale to another to realistically mimic the behavior of a living organism. As previously discussed, the possibility that a perturbation at the lowest scale (e.g., a random mutation or a mRNA intervention) can alter biological processes at larger scales is a mandatory feature for any reliable design of a digital twin.

\section*{Discussion and outlooks}

Innovative approaches for model integration within digital twins have a huge transformative potential for applications to precision medicine, enabling a synergy between generative modeling, advanced AI and machine learning techniques, and traditional biomedical insights. The fusion of these techniques, rather than the choice of a specific one, is expected to facilitate the development of new frameworks for multiscale modeling, which are pivotal in capturing the intricate dynamics of pathogenesis in humans. Through these frameworks, the overarching goal is to resolve the previously identified challenges, significantly enhancing the accuracy and clinical relevance of digital twins beyond inductive modelling via advanced statistics.

On the one hand, the integration of mechanistic models into digital twins also addresses the challenges of parameter indeterminacy and overfitting, which are prevalent in systems characterized by vast parameter spaces. By constraining these spaces, e.g. via the coarse-grained dynamics afforded by multiscale automata network models that synthesize large-scale data about biological mechanisms, digital twins not only gain in robustness and explainability but also offer a more reliable foundation for the simulation of therapeutic outcomes, thereby increasing their utility in clinical practice.

On the other hand, it is also worth discussing what is missing in current technologies and techniques developed for the same aim. For instance, a critical advantage of digital twins over
state-of-the-art non-computational models, such as organoids~\cite{corro2020brief}, is their capability to simulate complex, interdependent processes across multiple biological scales effectively, while providing explanatory and causal understanding and control at relatively small costs. 
While organoids can be engineered with all the power of modern synthetic biology~\cite{hofer2021engineering} to recapitulate features of the function and responses of complex biological mechanisms of the corresponding \emph{in vivo} target, they have important limitations.
Certainly, reproducibility is a major bottleneck~\cite{zhao2022organoids}, where digital twins can excel, especially if built under an open-source framework.
Additionally, organoids do not yet capture the entire physiological repertoire of cell types, even the behavior that is relevant for a particular disease. This means, for instance, that the response to drugs or other interventions need to be studied for organoids per se, separately from the \emph{in vivo} target. Related to this problem, is the relatively limited range of heterogeneity in response, which one needs to develop true personalized twins \cite{zhao2022organoids}.
Finally, while organoids are more direct analogues of biomolecular mechanisms, they cannot incorporate  simultaneously the multiple scales and historical information about patients, including the microbiome and exposome, which are major factors in complex diseases such as cancer, depression and many chronic diseases.
This is where the comprehensive multiscale network- and data-driven digital twin approach is particularly crucial.
Many complex diseases unfold across various multiomic sub-systems and exposome history.
Modular computational architectures that can synthesize and integrate multiple subsystems as separate network layers or agent-based models are well within the realm of possibility.
They might require a robust non-specialized digital twin, effectively integrating different specialized ones, to accommodate complex interactions and interdependency of biological and exposome processes. 
while non-specialized may not allow individual patient precision, such approach could still increase precision to specific cohorts rather than the whole population. 
For diseases with more circumscribed features, however, specialized digital twins might offer a targeted and streamlined alternative, allowing for precise intervention strategies and outcome predictions that might perform as well or better than those based on organoids.

Another remarkable advantage of digital twins is that they allow a scenario-based modeling approach for actionable interventions -- akin to strategies routinely used in epidemic modeling for policy decision-making~\cite{hofman2021integrating} -- that enhances their applicability and safety in clinical settings. This method avoids the standard pitfalls of an oracle-like predictive model by allowing for the exploration -- via direct simulation -- of multiple clinical scenarios, thereby providing a robust tool for decision support in personalized medicine. The integration with massive data sets about disease or treatment progressions, providing a reliable statistical samples that can be stratified to approximate the characterizing features of a patient, will be crucial to validate the output of models.

Therefore, the expected output of such a machinery, model-informed and data-driven, would not just be a yes/no decision about the adoption of a therapeutic strategy or an intervention, but a whole spectrum of alternatives where advantages and disadvantages in adopting each plausible strategy are outlined to inform human decision-making.







\clearpage
\bibliographystyle{naturemag}
\bibliography{biblio}

\paragraph{Author contributions.} M.D.D. designed the research. M.D.D., L.A., G.C, V.dA., B.dC., L.R., J.R, R.S and F.Z wrote the manuscript.

\paragraph{Competing interests.} The authors declare no competing interests.

\paragraph{Correspondence.} Correspondence should be sent to \url{manlio.dedomenico@unipd.it}

\end{document}